\documentclass[english]{article}
\usepackage[T1]{fontenc}
\usepackage[latin9]{inputenc}
\usepackage{geometry}
\usepackage{amsmath}
\usepackage{graphicx}

\makeatletter
\newcommand{\lyxaddress}[1]{
\par {\raggedright #1
\vspace{1.4em}
\noindent\par}
}

\makeatother

\usepackage{babel}
\begin{document}

\title{Average position of quantum walks with an arbitrary initial state\thanks{Project supported by the National Natural Science Foundation of China(No. 61504113) and the High-level Personnel Project of Xiamen University of Technology(No. YKJ13016R)}}

\author{Li Min\thanks{Corresponding author. E-mail: lm@xmut.edu.cn}, Cheng
ZaiJun, Wang LingJie, Huang HaiBo}

\maketitle

\lyxaddress{\begin{center}
School of Opto-Electronic and Communication Engineering, Fujian Provincial
Key Laboratory of Optoelectronic Technology and Devices, Xiamen University
of Technology, XiaMen, 361024, China
\par\end{center}}
\begin{abstract}
We investigated discrete time quantum walks with an arbitrary initial
state $\mid\Psi_{0}(\theta,\phi,\varphi)\rangle=\cos\theta e^{i\phi}\mid0L\rangle+\sin\theta e^{i\varphi}\mid0R\rangle$
with a U(2) coin $U(\alpha,\beta,\gamma)$. We discover that the average
position $\overline{x}=\max(\overline{x})\cos(\alpha+\gamma+\phi-\varphi)$,
with coin operator $U(\alpha,\pi/4,\gamma)$ and initial state $\mid\Phi_{0}(\pi/4,\phi,\varphi)\rangle=(e^{i\phi}\mid0L\rangle+e^{i\varphi}\mid0R\rangle)\sqrt{2}/2$.
If we set initial state and coin operator to $\mid\Phi_{0}\rangle(\theta,\pi/2,0)=i\cos\theta\mid0L\rangle+\sin\theta\mid0R\rangle)$
and coin operator $U(0,\pi/4,0)$, for $\alpha+\gamma+\phi-\varphi=\pi/2$,
we discover that $\overline{x}=-\max(\overline{x})\cos(2\theta).$
Last we verify the result above, and obtain the summarize properties
of quantum walks with an arbitrary state. We get that $\overline{x}(\theta,\phi,\varphi,\alpha,\beta,\gamma,t)=\cos2\theta*\overline{x}_{|0L\rangle}(\beta,t)+\sin2\theta*\cos(\alpha+\gamma+\phi-\varphi)*\overline{x}_{(\mid0L\rangle+\mid0R\rangle)\sqrt{2}/2}(\alpha=\gamma=0,\beta,t)$.
If the average positions $\overline{x}$ with initial state $|0L\rangle$
and $\mid\Psi_{0}\rangle=(\mid0L\rangle+\mid0R\rangle)\sqrt{2}/2$
and coin operator $U(0,\beta,0)$ are known, we can get the average
position result of quantum walks with an arbitrary initial state and
a U(2) coin operator.
\end{abstract}



\section{Introduction}

Quantum walks (QWs) were first introduced in 1993 \cite{Aharonov}
as the quantum version of classical random walks. QWs can be divided
into discrete time and continuous time \cite{Farhi-continue-time}
QWs. Both continuous time \cite{Childs-continue-universal} and discrete
time \cite{Lovett-discrete-universal,child2013} QWs can attain universal
quantum computation. QWs have found widespread applications in quantum
algorithms\cite{Childs2003,Shenvi2003,Childs2002,Childs2004,Ambainis,Ambainis2004}.
In addition, QWs in graph \cite{Aharonov-1}, on a line with a moving
boundary \cite{Kwek}, with multiple coins \cite{Brun-1}, decoherent
coins \cite{Brun} or a SU(2) coin \cite{Chandrashekar,average} have
been discussed.

Here we discuss the average position properties for QWs with an arbitrary
state and a U(2) coin.

\section{Discrete time quantum walks with an arbitrary initial state and a
U(2) coin operator}

The total Hilbert space for discrete time QWs is given by $\mathcal{H}\equiv\mathcal{H}_{P}\otimes\mathcal{H}_{C}$,
where $\mathcal{H}_{P}$ is spanned by the orthonormal states $\left\{ |x\rangle,x\in Z\right\} $
and $\mathcal{H}_{C}$ is the two-dimensional coin space spanned by
two orthonormal states $\mid L\rangle$ and $\mid R\rangle.$

In this paper, we discuss the QWs with an arbitrary initial state
and a U(2) coin operator. The probability distributions are the same
with a U(2) and a SU(2) coin operator\cite{average}. Here we use
a SU(2) matrix:

\begin{equation}
U(\alpha,\beta,\gamma)=\left(\begin{array}{cc}
e^{i\alpha}\cos\beta, & -e^{-i\gamma}\sin\beta\\
e^{i\gamma}\sin\beta, & e^{-i\alpha}\cos\beta
\end{array}\right)\label{eq:U(2)}
\end{equation}
as the coin operator. The particle movement operator is given by

\begin{equation}
S=\sum_{x}(|x-1\rangle\langle x|\otimes|L\rangle\langle L|+|x+1\rangle\langle x|\otimes|R\rangle\langle R|)
\end{equation}

After $t$ steps QWs, the final state can be written as:

\begin{equation}
\mid\Psi_{t}\rangle=\left[SU(\alpha,\beta,\gamma)\right]^{t}\mid\Psi_{0}\rangle
\end{equation}
where $\mid\Psi_{0}\rangle$ is an arbitrary initial state, is given
by:
\begin{equation}
\mid\Psi_{0}(\theta,\phi,\varphi)\rangle=\cos\theta e^{i\phi}\mid0L\rangle+\sin\theta e^{i\varphi}\mid0R\rangle.
\end{equation}

\section{Average position of quantum walks with an arbitrary initial state}

In this paper, we set the coin operator to $U(\alpha,\beta,\gamma)$,
and set the initial state to $\mid\Phi_{0}(\theta,\phi,\varphi)\rangle$,
then observe the change rule of average position with the changing
of parameters $\alpha$, $\gamma$, $\theta$, $\phi$ and $\varphi$.

\begin{figure}
\begin{centering}
\includegraphics[width=5in]{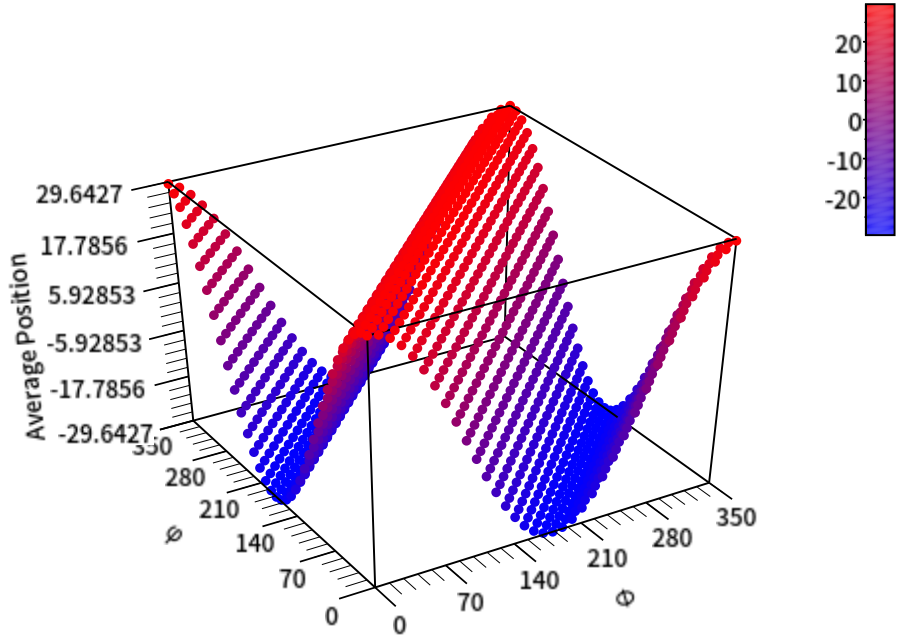}
\par\end{centering}

\protect\caption{(Color online) The average position $\overline{x}$ of quantum walks
after $t=100$ steps with SU(2) coin operator $U(0,\pi/4,0)$ and
initial state $\mid\Phi_{0}(\pi/4,\phi,\varphi)\rangle=\frac{\sqrt{2}}{2}(e^{i\phi}\mid0L\rangle+e^{i\varphi}\mid0R\rangle)$.}

\label{fig:1}
\end{figure}

\begin{figure}
\begin{centering}
\includegraphics[width=5in]{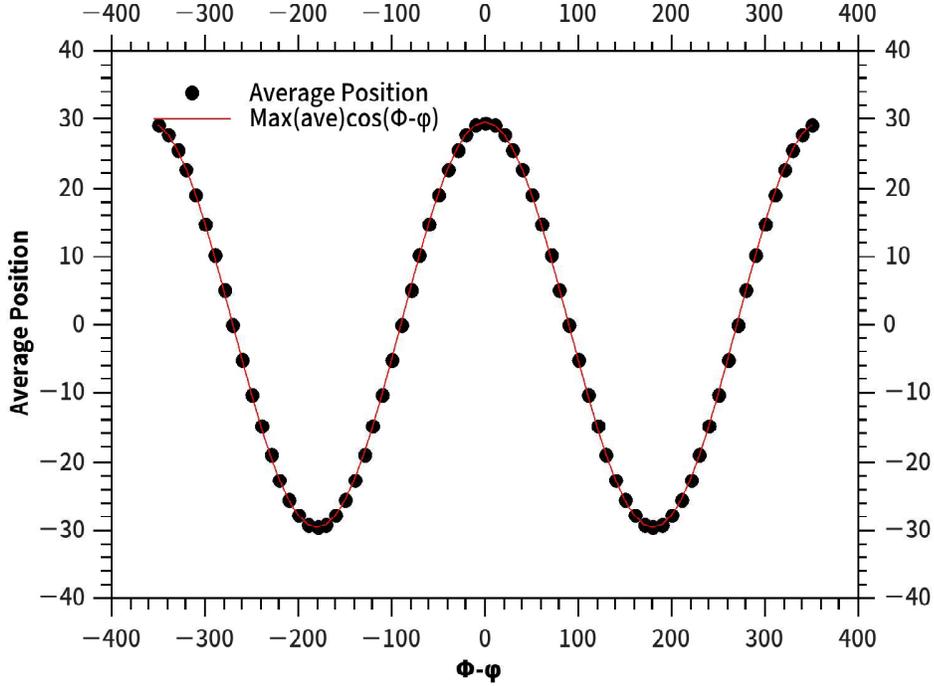}
\par\end{centering}

\protect\caption{(Color online) (black dot) The average position $\overline{x}$ with
$\phi-\varphi$ after $100$ steps quantum walk with SU(2) coin operator
$U(0,\pi/4,0)$ and initial state $\mid\Phi_{0}(\pi/4,\phi,\varphi)\rangle=\frac{\sqrt{2}}{2}(e^{i\phi}\mid0L\rangle+e^{i\varphi}\mid0R\rangle)$,
(red line) function $f(\phi-\varphi)=\max(\overline{x})\cos(\phi-\varphi)$. }

\label{fig:1-2}
\end{figure}

Figures \ref{fig:1} and \ref{fig:1-2} show the average position
$\overline{x}$ for QWs after 100 steps with a SU(2) coin $U(0,\pi/4,0)$,
while the initial state is $\mid\Phi_{0}(\pi/4,\phi,\varphi)\rangle=\frac{\sqrt{2}}{2}(e^{i\phi}\mid0L\rangle+e^{i\varphi}\mid0R\rangle)$.
From Fig. \ref{fig:1}, we can know that $\overline{x}$ only depends
on $\phi-\varphi$. Fig. \ref{fig:1-2} shows that the actual $\overline{x}$
exactly matches the function 
\begin{equation}
f(\phi-\varphi)=\max(\overline{x})\cos(\phi-\varphi).\label{eq:cos}
\end{equation}

\begin{figure}
\begin{centering}
\includegraphics[width=5in]{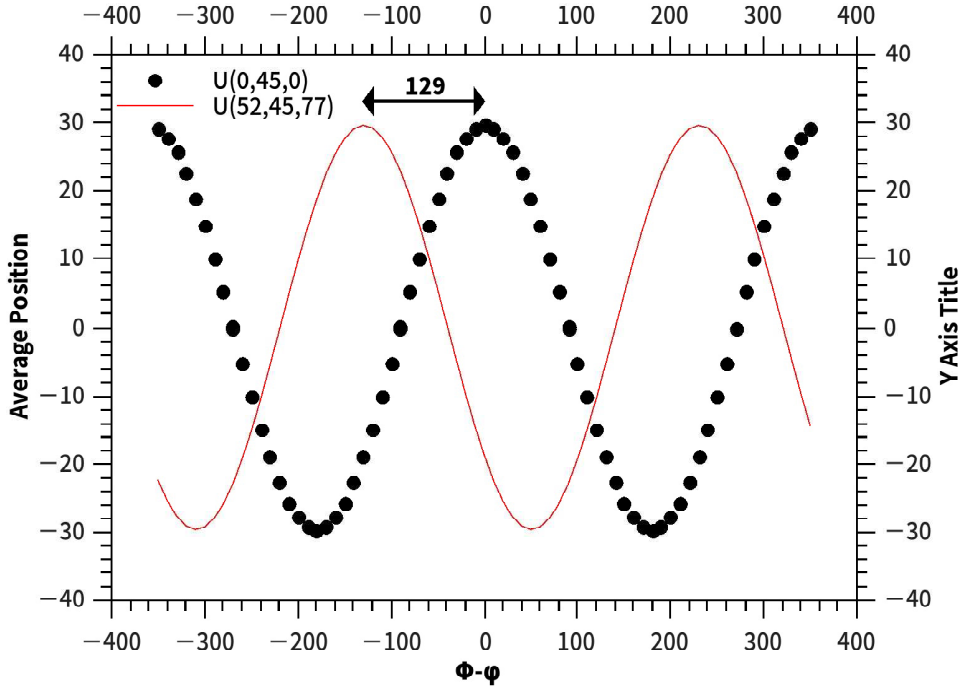}
\par\end{centering}

\protect\caption{(Color online)The average position of quantum walks with initial state
$\mid\Phi_{0}(\pi/4,\phi,\varphi)\rangle=\frac{\sqrt{2}}{2}(e^{i\phi}\mid0L\rangle+e^{i\varphi}\mid0R\rangle)$
after $100$ steps while the coin operators are $U(0,45^{\circ},0)$
(black dot) and $U(52^{\circ},45^{\circ},77^{\circ})$ (red line).
The angle shift between the two peak values is $129^{\circ}=52^{\circ}+77^{\circ}$.}

\label{fig:2}
\end{figure}

Fig. \ref{fig:2} shows that the average position of QWs after 100
steps with different coin operators, but the same initial state $\mid\Phi_{0}(\pi/4,\phi,\varphi)\rangle=\frac{\sqrt{2}}{2}(e^{i\phi}\mid0L\rangle+e^{i\varphi}\mid0R\rangle)$.
The two coin operators are $U(0,45^{\circ},0)$ (black dot) and $U(52^{\circ},45^{\circ},77^{\circ})$
(red line) respectively. From Fig. \ref{fig:2}, we can know that
the angle shift between the two peak values is $129^{\circ}=52^{\circ}+77^{\circ}$,
it means that the curve of red line is left shift $129^{\circ}$ to
the curve of black dot in Fig. \ref{fig:2}. Then we can conjecture
that the average position of QWs with coin operator $U(\alpha,\pi/4,\gamma)$
and initial state $\mid\Phi_{0}(\pi/4,\phi,\varphi)\rangle=\frac{\sqrt{2}}{2}(e^{i\phi}\mid0L\rangle+e^{i\varphi}\mid0R\rangle)$
can be given by:
\begin{equation}
\overline{x}=\max(\overline{x})\cos(\alpha+\gamma+\phi-\varphi).\label{eq:alpha+gamma}
\end{equation}

\begin{figure}
\begin{centering}
\includegraphics[width=5in]{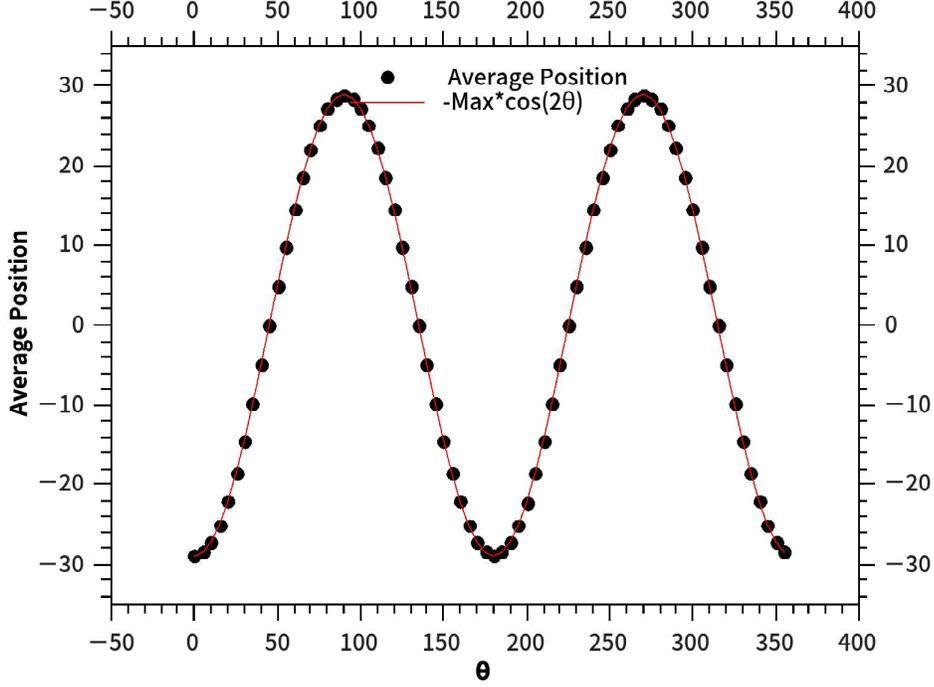}
\par\end{centering}

\protect\caption{(Color online) (black dot)The average position of quantum walks after
$100$ steps with the changing of initial state $\mid\Phi_{0}\rangle(\theta,\pi/2,0)$
and coin operator $U_{C}(0,\pi/4,0)$, (red line) $f(\theta)=-\max(\overline{x})\cos(2\theta)$. }

\label{fig:3}
\end{figure}

Fig. \ref{fig:3} shows the average position of QWs after $100$ steps
with the changing of initial state $\mid\Phi_{0}\rangle(\theta,\pi/2,0)=i\cos\theta\mid0L\rangle+\sin\theta\mid0R\rangle$
and coin operator $U(0,\pi/4,0)$, Fig. \ref{fig:3} shows that the
actual $\overline{x}$ exactly matches the function 
\begin{equation}
f(\theta)=-\max(\overline{x})\cos(2\theta).\label{eq:cos-1}
\end{equation}

\section{Proof in Mathematics}

From the Ref.  \cite{average}, we know that if the initial state
$\mid\Psi_{0}(\theta,\phi,\varphi)\rangle=\cos\theta e^{i\phi}\mid0L\rangle+\sin\theta e^{i\varphi}\mid0R\rangle$,
the coin operator
\[
U(\alpha,\beta,\gamma)=\left(\begin{array}{cc}
e^{i\alpha}\cos\beta, & -e^{-i\gamma}\sin\beta\\
e^{i\gamma}\sin\beta, & e^{-i\alpha}\cos\beta
\end{array}\right)
\]
 the probability $P^{L}(x)$ at state $|xL\rangle$ and $P^{R}(x)$
at state $|xR\rangle$ after $t$ steps of quantum walks can be given
by

\begin{equation}
\begin{cases}
P^{L}(x)=\cos^{2}\theta P_{\mid0L\rangle}^{L}(x)+\sin^{2}\theta P_{\mid0R\rangle}^{L}(x)-(e^{-i(\alpha+\gamma)}\cos\theta e^{-i\phi}\sin\theta e^{i\varphi}+e^{i(\alpha+\gamma)}\cos\theta e^{i\phi}\sin\theta e^{-i\varphi})G^{L}(\beta,x,t)\\
P^{R}(x)=\cos^{2}\theta P_{\mid0L\rangle}^{R}(x)+\sin^{2}\theta P_{\mid0R\rangle}^{R}(x)-(e^{-i(\alpha+\gamma)}\cos\theta e^{-i\phi}\sin\theta e^{i\varphi}+e^{i(\alpha+\gamma)}\cos\theta e^{i\phi}\sin\theta e^{-i\varphi})G^{R}(\beta,x,t)
\end{cases},\label{eq:PLR}
\end{equation}
where $P_{\mid0L\rangle}^{L}(x)$ and $P_{\mid0L\rangle}^{R}(x)$
are the probability at state $|xL\rangle$ and $|xR\rangle$ with
initial state $|0L\rangle$, $P_{\mid0R\rangle}^{L}(x)$ and $P_{\mid0R\rangle}^{R}(x)$
are the probability at state $|xL\rangle$ and $|xR\rangle$ with
initial state $|0R\rangle$, $G^{L}(\beta,x,t)$ and $G^{R}(\beta,x,t)$
are functions only depend on angle $\beta$, position $x$ and steps
$t$.

After $t$ steps, the average position of QWs $\overline{x}=\sum_{x=-t}^{t}x[P^{L}(\theta,\phi,\varphi,\alpha,\beta,\gamma,x,t)+P^{R}(\theta,\phi,\varphi,\alpha,\beta,\gamma,x,t)]$.

From Ref.  \cite{average}, we know that $P_{\mid0j\rangle}^{i}(\alpha,\beta,\gamma,x,t)=P_{\mid0j\rangle}^{i}(\beta,x,t),i,j\in\{L,R\}$
are irrelevant to the parameters $\alpha$ and $\gamma$, and $P_{\mid0L\rangle}^{R}(\beta,x,t)=P_{\mid0R\rangle}^{L}(\beta,-x,t)$,
$P_{\mid0L\rangle}^{L}(\beta,x,t)=P_{\mid0R\rangle}^{R}(\beta,-x,t)$.

Then we can get that

\begin{equation}
\begin{cases}
\sum_{x=-t}^{t}x[\cos^{2}\theta P_{\mid0L\rangle}^{L}(\beta,x,t)+\sin^{2}\theta P_{\mid0R\rangle}^{R}(\beta,x,t)]=\cos2\theta\sum_{x=-t}^{t}x[P_{\mid0L\rangle}^{L}(\beta,x,t)]\\
\sum_{x=-t}^{t}x[\cos^{2}\theta P_{\mid0L\rangle}^{R}(\beta,x,t)+\sin^{2}\theta P_{\mid0L\rangle}^{R}(\beta,x,t)]=\cos2\theta\sum_{x=-t}^{t}x[P_{\mid0L\rangle}^{R}(\beta,x,t)]
\end{cases}\label{eq:a10}
\end{equation}
and 
\begin{equation}
(e^{-i(\alpha+\gamma)}\cos\theta e^{-i\phi}\sin\theta e^{i\varphi}+e^{i(\alpha+\gamma)}\cos\theta e^{i\phi}\sin\theta e^{-i\varphi}=\sin2\theta*\cos(\alpha+\gamma+\phi-\varphi)\label{eq:a11}
\end{equation}

Using Eq. \ref{eq:a10} and \ref{eq:a11}, we can get the average
position 
\begin{equation}
\overline{x}=\cos2\theta*\overline{x}_{\mid0L\rangle}(\beta,t)+\sin2\theta*\cos(\alpha+\gamma+\phi-\varphi)*G(\beta,t)\label{eq:G}
\end{equation}
 where $G(\beta,t)=-\sum_{-t}^{t}x[G^{L}(\beta,x,t)+G^{R}(\beta,x,t)]$.

If we set $\theta=\pi/4$, $\alpha+\gamma+\phi-\varphi=0$, for easily,
$\alpha=\gamma=\phi=\varphi=0$, the coin operator is $U(0,\pi/4,0)=\frac{\sqrt{2}}{2}\left(\begin{array}{cc}
1, & -1\\
1, & 1
\end{array}\right)$, and the initial state is $\mid\Psi_{0}\rangle=\frac{\sqrt{2}}{2}(\mid0L\rangle+\mid0R\rangle)$,
we can get $G(\beta,t)=\overline{x}(\theta=\pi/4,\alpha=\gamma=\phi=\varphi=0,\beta,t)$.

Then Eq. \ref{eq:G} can be rewritten as 

\begin{equation}
\overline{x}(\theta,\phi,\varphi,\alpha,\beta,\gamma,t)=\cos2\theta*\overline{x}_{|0L\rangle}(\beta,t)+\sin2\theta*\cos(\alpha+\gamma+\phi-\varphi)*\overline{x}_{\sqrt{2}(\mid0L\rangle+\mid0R\rangle)/2}(\alpha=\gamma=0,\beta,t)\label{eq:ave}
\end{equation}
where $|0L\rangle$ and $\sqrt{2}(\mid0L\rangle+\mid0R\rangle)/2$
indicate the different initial states of QWs.

Using Eq. \ref{eq:ave}, we can get that $\overline{x}(\pi/4,\phi,\varphi,\alpha,\pi/4,\gamma,t)=\cos(\alpha+\gamma+\phi-\varphi)*\overline{x}_{\sqrt{2}(\mid0L\rangle+\mid0R\rangle)/2}(\alpha=\gamma=0,\beta,t)$,
this equation proves the results in Fig. \ref{fig:1}, \ref{fig:1-2}
and \ref{fig:2}. 

The same as above, we can get that $\overline{x}(\theta,\pi/2,0,0,\beta,0,t)=\cos2\theta*\overline{x}_{|0L\rangle}(\beta,t)$,
this equation proves the result in Fig. \ref{fig:3}. If we set $\overline{x}_{|0L\rangle}(\beta,t)=A$,
$\cos(\alpha+\gamma+\phi-\varphi)*\overline{x}_{\sqrt{2}(\mid0L\rangle+\mid0R\rangle)/2}(\alpha=\gamma=0,\beta,t)=B$,
Eq. \ref{eq:ave} can be rewritten as: 
\begin{equation}
\overline{x}(\theta,\phi,\varphi,\alpha,\beta,\gamma,t)=\sqrt{A^{2}+B^{2}}\cos(2\theta-\omega)
\end{equation}
 where $\cos\omega=\frac{A}{\sqrt{A^{2}+B^{2}}}$ is irrelevant to
the parameter $\theta$. The average position curve with $\theta$
still match the function of $\max(\overline{x})\cos2\theta$, but
right shift $\omega/2$.

If we set the initial state to $|\Phi_{0}(\pi/4,0,\pi/2)\rangle=1/\sqrt{2}(|0L\rangle+i|0R\rangle)$,
using Eq. \ref{eq:ave} we can get that $\overline{x}(\pi/4,0,\pi/2,\alpha,\beta,\gamma,t)=\sin(\alpha+\gamma)*\overline{x}_{\sqrt{2}(\mid0L\rangle+\mid0R\rangle)/2}(\alpha=\gamma=0,\beta,t)$,
this equation also proves the result in Ref. \cite{average}.

For any $\alpha+\gamma+\phi-\varphi=0$ , $\overline{x}(\theta=\pi/4,\alpha+\gamma+\phi-\varphi=0,\beta,t)=\overline{x}(\pi/4,0,0,0,\beta,0,t)$.

Eq. \ref{eq:ave} indicates the change rule of the average position
in QWs with the parameters $\theta$, $\phi$, $\varphi$, $\alpha$
and $\gamma$. If we know average positions $\overline{x}$ with coin
operator $U(0,\beta,0)$, while the initial states are $|0L\rangle$
and $\frac{\sqrt{2}}{2}(\mid0L\rangle+\mid0R\rangle)$, we can get
the average position result for an arbitrary state using Eq. \ref{eq:ave}.

\section{Conclusions}

In this paper, we discussed the average position property in QWs with
an arbitrary initial state $\mid\Psi_{0}\rangle(\theta,\phi,\varphi)=\cos\theta e^{i\phi}\mid0L\rangle+\sin\theta e^{i\varphi}\mid0R\rangle$
and a $U(2)=U(\alpha,\beta,\gamma)$ coin operator. Firstly, we set
the coin operator and initial state to $U(0,\pi/4,0)$ and $\mid\Phi_{0}(\pi/4,\phi,\varphi)\rangle$
respectively, we get that $\overline{x}=\max(\overline{x})\cos(\phi-\varphi)$.
Secondly, if we left the initial state unchanged, and set the coin
operator to $U(\alpha,\pi/4,\gamma)$, we get that $\overline{x}=\max(\overline{x})\cos(\alpha+\gamma+\phi-\varphi).$
Thirdly, we set the coin operator and initial state to $U(0,\pi/4,0)$
and $\mid\Phi_{0}(\theta,\pi/2,0)\rangle$ respectively, we get that
$\overline{x}=-\max(\overline{x})\cos(2\theta).$ Lastly, we poof
the above result in mathematics, we get that the average position
$\overline{x}(\theta,\phi,\varphi,\alpha,\beta,\gamma,t)=\cos2\theta\overline{x}_{|0L\rangle}(\beta,t)+\sin2\theta*\cos(\alpha+\gamma+\phi-\varphi)\overline{x}_{\sqrt{2}(\mid0L\rangle+\mid0R\rangle)/2}(\alpha=\gamma=0,\beta,t)$.
If we know average positions $\overline{x}$ with coin operator $U(0,\beta,0)$,
while the initial states are $|0L\rangle$ and $\mid\Psi_{0}\rangle=\frac{\sqrt{2}}{2}(\mid0L\rangle+\mid0R\rangle)$,
we can get the average position result of quantum walks for an arbitrary
state and a U(2) coin operator.

\end{document}